\documentclass{ws-ijmpe}
\usepackage[super,compress]{cite}
\begin{document}

\markboth{Xu Cao}{Disentangle the nature of resonances in coupled-channel models}

\catchline{}{}{}{}{}

\title{Disentangle the nature of resonances in coupled-channel models}

\author{\footnotesize XU CAO}

\address{Institute of Modern Physics, Chinese Academy of Sciences, Lanzhou 730000, China\\
Institut f\"{u}r Theoretische Physik, Universit\"{a}t Giessen, D-35392 Giessen, Germany\\
State Key Laboratory of Theoretical Physics, Institute of Theoretical Physics,\\ Chinese
Academy of Sciences, Beijing 100190, China\\
caoxu@impcas.ac.cn}

\maketitle

\begin{history}
\received{Day Month Year}
\revised{Day Month Year}
\end{history}

\begin{abstract}
  We present several possible hadronic states found in coupled-channel models within the on-shell approximation. The interaction potential is constructed as a sum of the tree-level Feynman diagrams calculated with the effective Lagrangians. Based on the recent empirical data, we illustrate the possible existence of several baryonic and mesonic states with definite quantum numbers in the model. We give their properties for the purpose of further study and discuss the potential of finding them in future experiments.
\end{abstract}

\keywords{Partial wave analysis; coupled-channel models; exotic state.}

\ccode{PACS numbers: 11.80.Et, 13.20.Gd, 14.20.Gk, 14.40.Rt}


\section{Introduction} \label{sec:intro}

Nowadays, the Quantum Chromodynamics (QCD) has not yet been numerically resolved at the range of low energies, if not mentioning the lattice QCD for the moment. Instead, in order to reveal the nature of hadronic states, a lot of sophisticated phenomenological models are developed to separate the resonance from the overlapped background and coupled-channel effect. In this energy range of the non-perturbative Quantum Chromodynamics, the coupled-channel effect makes the analysis complicated and partial wave analysis (PWA) seems to be an useful tool of analyzing the baryonic and mesonic states. On the experimental side, the scattering of stable particles and the hadronic decay of heavy quarkonium are usually used to extract empirical information of resonances. The typical examples are the $\pi N$, $\gamma N$~\cite{Penner2002,caoKSigma,LeePhysRe,KamanoCC,Kamano2pi}, $e N$ and $p N$ reactions~\cite{caoCPL2008,caoetap2008,caophi2009,caoCPC2009,caotwopi,caoNPA2011,caoIJMPA,caoAIP} and the decay of the charmonium states, e.g. $J/\psi$, $\psi '$ and $\chi_{cJ}$~\cite{ZouEPJA2003,ZouPRC2003,ZouEPJA2005}.

The analyses of hadronic decay of heavy quarkonium are simplified by the isospin filter which only allows the contribution of isospin zero. A lot of progress has been made in this direction in recent years~\cite{BEShadron3,BEShadron4}. The states with isospin one should be studied in other reactions, but they are usually convoluted with isospin zero components. In the $p N$ reactions~\cite{caoCPL2008,caoetap2008,caophi2009,caoCPC2009,caotwopi,caoNPA2011,caoIJMPA,caoAIP}, the coupled-channel effects are complicated due to three or more particles in final states so the study is still in the initial stage. For the photo-induced reactions, though the non-resonant background is large due to the enhancement of Born terms resulting from the gauge invariance~\cite{caoKSigma}, the unprecedented development of coupled-channel models has reliably dissociated the resonances from background and investigated the properties of many states by treating the amount of $\pi N$ and $\gamma N$ data on the equal footing. One kind of the coupled-channel models is proceeded from the Bethe-Salpeter equation in the on-shell approximation~\cite{Penner2002,caoKSigma}:
\begin{equation}
T = V +VGT = \frac{V}{1 - VG},
\label{BSEqfull}
\end{equation}
where $T$ stands for the total transition amplitudes and $V$ is the interaction kernel in the tree level, respectively. The propagators matrix $G$ is diagonal. On the basis of fundamental scattering theory, the wave function could be expanded in the eigenstates of the angular momentum. As a result, the transition amplitudes in the above equation could be decomposed into series of partial waves with definite quantum numbers. Then the scattering equation can be solved for each partial wave separately. This PWA are proved to be a powerful tool to compare different models and experimental approaches with each other. The partial waves in different models give a direct comparison in the details of various approaches and serve as a guideline for future improvements. In addition, PWA can be used to decompose the measured sample of events into the constituent contributions of partial waves, so give a straight link to the model calculation and experiment. In the following text we would like to show that several baryonic and mesonic states could be found in this scheme based on the experimental data.

\section{Baryons in the photo-induced reactions} \label{sec:baryon}

The Bethe-Salpeter equation Eq.(\ref{BSEqfull}) is extraordinarily complicated because its denominator involves four-dimensional loop integration. In order to reduce this daunting integration task and numerically solve the Eq.(\ref{BSEqfull}) in meson-baryon interaction channels, the approximation is unavoidable. The propagator $G$ in Eq.(\ref{BSEqfull}) could be split into two constituents containing the real and imaginary parts. In the so-called K-matrix approximations, the real part of the propagator is neglected. This is justified considering that the real part of the intermediate loop integrals only results in a non-observable renormalization of the coupling constants and masses of the involved particles. The validity of this approximation has been extensively tested and explored in the literature~\cite{Penner2002,caoKSigma}. After some algebraic manipulation, the scattering $T$-matrix is reduced to the set of equations for each partial wave~\cite{Penner2002,caoKSigma}:
\begin{equation}
T^{J\pm,I}_{fi} = \left[\frac{ K^{J\pm,I}}{1-iK^{J\pm,I}}\right]_{fi}\,,
\label{BSEqK}
\end{equation}
where $J\pm$ and $I$ are total spin, parity and isospin of the initial and final states. Below the center of mass energy 2000 MeV, the $f,i =$ $ \gamma N$, $\pi N$, $2\pi N$, $\eta N$, $\omega N$, $K\Lambda$ and $K\Sigma$ channels are opened and should be included into the model. The kernel $K = V$ could be dynamically built as the $s$, $u$, and $t-$ channels with the effective interaction Lagrangians respecting chiral symmetry in low-energy regime\cite{Penner2002}. Recently we analyze the current data of $K\Sigma$ photoproduction in this framework, and the parameters are well constrained by the new data together with the previous $\pi N$ and $\gamma N$ partial waves~\cite{caoKSigma}. Here we would like to address the indication of a $\Delta^*$ resonance around 2000 MeV, namely the $F_{35}(2000)$ state.

\begin{figure}[th]
\centerline{\psfig{file=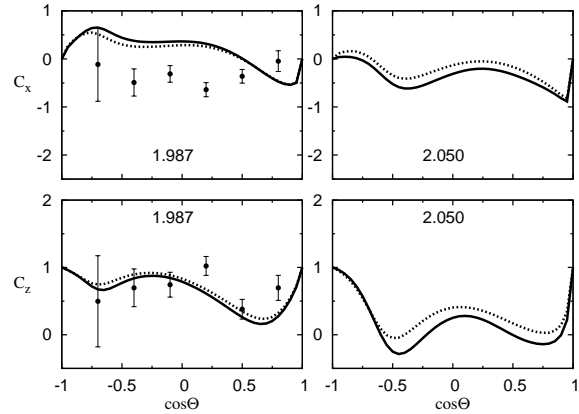,width=8.cm}}
\caption{The spin transfer coefficient $C_{x}$ and $C_{z}$ in $\gamma p \to K^+ \Sigma^0$ reaction. The dashed and solid lines are the full model calculation and the model calculation with the $F_{35}(2000)$ turned off, respectively. Data are taken from CLAS~\cite{CLAS07CxCz}. The numeric values label the center of mass energies in unit of GeV.}  \label{Fig:PoKS}
\end{figure}

In the current version of the compilation of Particle Data Group~\cite{pdg2012}, the $F_{35}(2000)$ is only rating as a two-star resonance, which is meaning that it has little experimental evidence up to now. Its mass and width are found to be 2160 and 313 MeV in our analysis, respectively. These values should be compared to 2015(24) and 500(52) MeV in KSU survey~\cite{KSU2012}, the only PWA group presently including this state. Both results find that its major decay channel is the $2\pi N$, whose partial width is bigger than 90\%. After adding the $F_{35}(2000)$ into our model, it improves the high energy tail of the partial waves of elastic $\pi N$ channels. So it seems to be of non-resonant nature and play the role of background in the $\pi N$ reactions. This is telling us why this state is hard to present clearly in the $\pi N$ collisions. However, after shutting off this state, the $\chi^2$ increases a value of 2.3 in the $\gamma p \to K\Sigma$ channel. Though its contribution to the total cross section is tiny, the role in the polarization observables is clearly seen, as depicted in Fig.~\ref{Fig:PoKS}. Above 2.0 GeV it is shown that the $F_{35}(2000)$ would be more obvious. Other resonances with different quantum numbers, e.g. $S_{11}$, $S_{31}$, $P_{11}$ and $P_{31}$ resonance with Breit-Wigner mass varying from 1700 Mev to 2000 MeV are tried to add to the model but we do not find any improvement in the description of the current $K\Sigma$ photoproduction data.

As a matter of fact, the $F_{35}(2000)$ is found to be important in describing the differential cross section of the $\Sigma$(1385) photoproduciton from proton within a Regge-plus-resonance approach~\cite{He2013}. The determined helicity amplitudes $A_{1/2} = -10$  GeV$^{-1/2}$ and $A_{1/2} = -28$ GeV$^{-1/2}$ should be compared to our fitted values 18 GeV$^{-1/2}$ and -23 GeV$^{-1/2}$, respectively. The corresponding values of KSU survey~\cite{KSU2012} are -16(81) GeV$^{-1/2}$ and 158(32) GeV$^{-1/2}$. Though these values have large uncertainties, they consistently indicate the relative big radiative decay width of $F_{35}(2000)$, so the photoproductions are expected to be suitable for studying its properties. In order to resolve unambiguously the signal of $F_{35}(2000)$, it is necessary to extend the coupled-channel model to higher energies, where the $K\Sigma$(1385), $K^*\Sigma$ and $K^*\Lambda$ channels are opening. It would be also meaningful to explore the $3 \pi N$ channel because strong coupling of $F_{35}(2000)$ to $\pi\pi\Delta$ is found~\cite{XiePRC2011}. It is meaningful to perform this kind of analysis considering the increasing experimental database in this energy range.

\section{Hunting for exotic mesons} \label{sec:meson}

\begin{table}[pt]
\tbl{Pole positions of non-strangeness states in coupled-channel model with chiral Lagrangians.\label{tab:poles}}
{\begin{tabular}{@{}c||c|c|c@{}} \toprule
  &I$^G(J^{PC})$& RE($\sqrt{s}$) (MeV)& IM($\sqrt{s}$) (MeV) \\ \colrule
VV&$0^+(0^{++})$& 10650.2$\pm$1.9\hphantom{0} & 11$\pm7$   \\
  &$0^-(1^{+-})$& 10471.1$\pm$42.3\hphantom{0}& 1$\pm1$   \\
  &             & 10650.2$\pm$1.5\hphantom{0}    & 127$\pm$16 \\
  &$0^+(2^{++})$& 10650.2$\pm$1.7\hphantom{0} & 13$\pm$8   \\
  &$1^-(0^{++})$& 10650.2$\pm$1.6\hphantom{0} & 7$\pm$5   \\
  &$1^+(1^{+-})$& 10650.2$^{(a)}$             & 5$\pm$4   \\
  &$1^-(2^{++})$& 10650.2$\pm$1.6\hphantom{0} & 7$\pm$3   \\ \colrule
VP&1$^+(1^{+-})$& 10604.2$^{(b)}$             & -6$\pm4$  \\
  &1$^-(1^{++})$& ---                         & --- \\
  &0$^+(1^{++})$& 10604.2$\pm$1.7\hphantom{0} & -95$\pm$17 \\
  &             & 10781.0$\pm$1.3\hphantom{0} & -118$\pm12$ \\
  &0$^-(1^{+-})$& 10781.7$\pm$1.3\hphantom{0} & -2$\pm$1  \\ \colrule
PP&0$^+(0^{++})$& 10732.8$\pm$1.6\hphantom{0} & -6$\pm$4  \\
  &1$^-(0^{++})$& 10558.0$\pm$1.1\hphantom{0} & -7$\pm$4 \\ \botrule
\end{tabular}}
\begin{tabnote}
They are set as input~\cite{Zb2011}:
\end{tabnote}
\begin{tabfootnote}
\tabmark{a} $Z_{b}(10650)$ with M=10652.2$\pm$1.5 MeV and $\Gamma$=11.5$\pm$ 2.2 MeV\\
\tabmark{b} $Z_{b}(10610)$ with M=10607.2$\pm$2.0 MeV and $\Gamma$=18.4$\pm$ 2.4 MeV
\end{tabfootnote}
\end{table}

In the energy range below 2.0 GeV, the exotic mesons usually have broad widths and overlap with other non-exotic states. So it is very hard to clearly establish them though a lot of efforts have been devoted to the PWA within isobar models~\cite{MeyerPRC2010}. But in the charm and bottom sector, the widths of mesons are limited by flavor changing weak interactions so the possibility of discovery of exotic states are larger. A typical example is the charged Bottomonium-like states, the $Z_{b}(10610)$ and $Z_{b}(10650)$, found in the decay of the Bottomonium states~\cite{Zb2011}. Their masses are close to $B^*\bar{B}$ and $B^*\bar{B}^*$ threshold, respectively and the observed decay channel requires a $b\bar{b}d\bar{u}$ quark flavor structure component. If they are confirmed, then we expect other states in the bottom sector in the coupled-channel model, as hinted by the heavy quark flavor symmetry~\cite{GuoHQS2013}. As a simple illustration, we construct the kernel $V$ from chiral Lagrangians within the framework of the hidden gauge formalism, similar to the case in charm sector. Then the detailed formalism of kernel $V$, decomposed in terms of partial waves, could be adopted from Ref.~\cite{Roca2005,Geng2009,MolinaPRD2009,Gamermann}. Following this prescription, the propagator of the exchanged meson is proportional to the inverse of its squared mass, so heavy meson exchange is suppressed by a factor $\gamma=(m_L/ m_H)^2$, which is also the scale of the SU(4) symmetry broken.

The propagator matrix $G$ can be calculated as two-meson loop function using dimensional regularization~\cite{MolinaPRD2009}:
\begin{eqnarray}
G_l &=& \frac{1}{16\pi ^2}\biggr( \alpha_l+Log\frac{m_l^2}{\mu ^2}+\frac{M_l^2-m_l^2+s}{2s}
  Log\frac{M_l^2}{ m_l^2}+\nonumber \\
& &\frac{p}{\sqrt{s}}\Big( Log\frac{s-M_l^2+m_l^2+2p\sqrt{s}}{ -s+M_l^2-m_l^2+
  2p\sqrt{s}} \nonumber\\
& & +Log\frac{s+M_l^2-m_l^2+2p\sqrt{s}}{-s-M_l^2+m_l^2+  2p\sqrt{s}}\Big)\biggr),
  \label{loopf}
\end{eqnarray}
where $M_l$ and $m_l$ are the masses of the intermediate two mesons and $p$ is the corresponding three-momentum in the center of mass frame. The cut-off $\mu$ is set to 2.1 GeV and the $\alpha_l$ is the subtraction constant. Unitarity is ensured by the equation $Im(G_l)=- p / 8\pi\sqrt{s}$, so the states are present as poles in the complex plane of the full $T$-matrix in Eq.(\ref{BSEqfull}). The two-meson channels are included as:

I. Vector-Vector (VV) interaction: $B^*\bar{B}^*$, $B_s^*\bar{B}_s^*$, $K^*\bar{K}^*$, $\rho\rho$, $\omega\omega$, $\phi\phi$, $\omega\phi$, $\Upsilon(1S)\Upsilon(1S)$, $\Upsilon(1S)\omega$, $\Upsilon(1S)\phi$;

II. Vector-Pseudoscalar (VP) interaction: $B^*\bar{B}$, $B_s^*\bar{B}_s$, $K^*\bar{K}$, $\rho\pi$, $\phi\eta(')$, $\omega\eta(')$, $\Upsilon(1S)\eta(')$, $\Upsilon(1S)\eta_b$, $\eta_b \omega$, $\eta_b \phi$;

III. Pseudoscalar-Pseudoscalar (PP) interaction: $B\bar{B}$, $B_s\bar{B}_s$, $K\bar{K}$, $\pi\pi$, $\eta(')\eta(')$, $\eta_b\eta_b$, $\eta_b\eta(')$.

The masses of above mesons are adopted from Particle Data Group~\cite{pdg2012}. The values of parameters required to be the input of this coupled-channel model are chosen to be the characterizing values~\cite{pdg2012,Lattice2007}:

I. $m_L = 800.0$~MeV and $m_H = 5370.0$~MeV for the suppression factor $\gamma$;

II. The decay constant: $f_\pi = 93.0$~MeV, $f_B = 195.0/\sqrt{2}$~MeV, $f_{Bs} = 250.0/\sqrt{2}$~MeV, and $f_{\eta_b} = 801.0/\sqrt{2}$~MeV.

If the $Z_{b}(10610)$ and $Z_{b}(10650)$ are assumed to be dynamically generated states with I$^G(J^{PC}) = 1^+(1^{+-})$, close to $B^*\bar{B}$ and $B^*\bar{B}^*$ thresholds respectively, then the free parameter $\alpha_l$ could be determined to be -3.12$\pm$0.08 for the channels with two heavy vector mesons and -1.65$\pm$0.05 for other cases. The uncertainties are estimated from the experimental errors of the masses and widths of the two $Z_{b}$ states. Then the states of other partial waves could be predicted in Tab.\ref{tab:poles}. As can be seen, the states close to the $B^*\bar{B}^*$ threshold are expected to be present in all quantum numbers in the VV channel. These states are very narrow except the $0^-(1^{+-})$ state with around 200 MeV width. We also predict another narrow $0^-(1^{+-})$ state located at about 10471.1 MeV, but its mass has relatively big uncertainty.  In the PP channel, a 0$^+(0^{++})$ state with the pole at around 10732.8 MeV and a 1$^-(0^{++})$ state at about 10558.0 MeV are expected, whose widths are both small.

In the Lippmann-Schwinger equation respecting heavy quark flavor symmetry, a 0$^+(1^{++})$ $B^*\bar{B}$ bound state with a mass about 10580 MeV is predicted~\cite{GuoHQS2013}, which should be compared to two broad states with the poles respectively at 10604.2 MeV and 10781.0 MeV in our coupled-channel model. In addition, our model predicts a narrow 0$^-(1^{+-})$ state with the pole at 10781.0 MeV. So it is possible to distinguish these two models in the VP channel by experiments.

In Ref.~\cite{Hosaka2012} the coupled-channel Schr\"{o}dinger equation is solved numerically with the meson exchange potentials. They predict several possible bound and resonant states after reproducing the $Z_b$(10610) and $Z_b$(10650). A similar framework addresses the role of $S$-$D$ mixing~\cite{ZFSun2011,NLi2013}. These works find that $\pi$-meson exchange is important~\cite{Hosaka2012,ZFSun2011,NLi2013,OsetPRD2013,Sudoh2009,Hosaka2011}. However, in Ref.~\cite{GuoHQS2013} it is shown that one pion exchange only slightly changes the central value of the calculated mass. In our model, we assume the vector-meson dominance. So it is still an open question as to which interaction, $\pi$- or vector-meson exchange, is dominant. The predicted masses of the states in different models would be tested by future experiments.

It is worth mentioning that while the broad states would be completely immersed into background so not reachable in the high energy machine, the predicted narrow states could emerge in future experiment. We suggest the experimentalists to search for these narrow states in the photo-induced reactions, i.e. $\gamma N \to N Z_b$ reactions. The maximum total cross section of Charmonium-like mesons in the photo-induced reactions is estimated to be around $\sigma_{Zc} \lesssim 10nb$ with the assumption of dominance of $t$-channel vector-meson exchange~\cite{HePRD2009,LinPRD2014}. If the width of Bottomonium-like mesons decaying to $\Upsilon(1S)V$ or $\eta_b V$ channel is at the level of MeV, similar to that of Charmonium-like states, the production cross sections of $\gamma N \to N Z_b$ reactions could be roughly estimated to be at the level:
\begin{eqnarray}
\sigma_{Zb} = \sigma_{Zc} \frac{s_c (k_c^{\gamma} f_{\eta_c})^2}{s_b (k_b^{\gamma} f_{\eta_b})^2} \lesssim 0.5nb
\end{eqnarray}
where $s_b$ and $k_b^{\gamma}$ are invariant mass squared and photon energy in the center of mass system. The factors $s_b (k_b^{\gamma})^2$ and $s_c (k_c^{\gamma})^2$ could be understood from phase spaces and the decay constants $f_{\eta_b}$ and $f_{\eta_c}=420.0/\sqrt{2}$~MeV~\cite{MolinaPRD2009} from the production amplitudes. Though production cross sections are small, the background is clean at this energy range and under control by the Pomeron exchange, leading to the possibility of finding these states, especially the exotic states with $1^{+-}$ quantum numbers.

\section{Remarks and Conclusion} \label{sec:con}

The rapid advancement of the coupled-channel model and the wide application of the PWA tools have promoted the recent progress of hadronic spectra. The coupled-channel model with effective or chiral Lagrangians could be used to predict several hadronic states based on the current empirical information. In this paper we have addressed the evidence of several baryonic and mesonic states with definite quantum numbers in the coupled-channel models. We point out the $K\Sigma$(1385) photoproduction data would offer a good chance to confirm the $F_{35}(2000)$, a state not yet well established in the current $K\Sigma$ photoprodunction data. We have also shown the coupled-channel model on the basis of the known states could be used to predict the Bottomonium-like mesons. Though the production cross section is estimated to be small, the photo-induced reactions in future high luminosity accelerator seem to have the potential for further studying these states and disentangle the nature of resonances in different models.

\section*{Acknowledgements}

We would like to thank Prof. H. Lenke and Dr. V. Shklyar for collaboration in the work of Sec.~\ref{sec:baryon}. Useful discussions with Prof. B. S. Zou, Prof. X. R. Chen and Dr. J. J. Xie are gratefully acknowledged. This work was supported by the National Natural Science Foundation of China (Grant No.11347146, No.11275235 and No. 11175220) and the One Hundred Person Project (Grant No. Y101020BR0).

\end{document}